\documentclass[11pt]{article}

\usepackage{graphicx} 
\usepackage[utf8]{inputenc}
\usepackage{amsmath}
\usepackage{amsfonts}
\usepackage{caption}
\usepackage{graphicx}
\usepackage{amssymb}
\usepackage{dirtytalk}
\usepackage[dvipsnames,table,xcdraw]{xcolor}
\usepackage{comment}
\usepackage{enumerate}   
\usepackage{bbm}
\usepackage{amsthm}
\usepackage{thmtools}
\usepackage[numbers,sort]{natbib}
\usepackage{comment}
\usepackage{algorithm}
\usepackage{algpseudocode}
\usepackage[margin=1in]{geometry}
\usepackage{pgfplots}
\usepackage{crossreftools}
\usepackage{multirow}
\usepackage{bm}
\usepackage{empheq}
\usepackage{microtype}
\usepackage{xurl}
\usepackage{hyperref}
\usepackage{cleveref} 
\usepackage{tabularx}
\usepackage{booktabs}

\pgfplotsset{compat=1.18}
\hypersetup{
    colorlinks=true,
    linkcolor=Red,
    citecolor=Cyan,
    urlcolor=Magenta
}

\makeatletter
\def\thmt@refnamewithcomma #1#2#3,#4,#5\@nil{%
  \@xa\def\csname\thmt@envname #1utorefname\endcsname{#3}%
  \ifcsname #2refname\endcsname
    \csname #2refname\expandafter\endcsname\expandafter{\thmt@envname}{#3}{#4}%
  \fi
}
\makeatother

\newtheorem{theorem}{Theorem}[section]

\newtheorem{lemma}[theorem]{Lemma}

\newtheorem{corollary}[theorem]{Corollary}
\newtheorem{proposition}[theorem]{Proposition}
\newtheorem{inftheorem}{Informal Theorem}

\newcounter{algorithmicinstance}
\AddToHook{env/algorithmic/begin}{\stepcounter{algorithmicinstance}}
\expandafter\def\csname theHALG@line\endcsname{\arabic{algorithmicinstance}.\arabic{ALG@line}}
\makeatletter
\crefalias{ALG@line}{line}
\AddToHook{env/algorithmic/begin}{%
    \let\algorithmiclabel\label
    \def\label{%
        \def\@currentcounter{ALG@line}%
        \protected@edef\cref@currentlabel{[line][\arabic{ALG@line}][]\theALG@line}%
        \algorithmiclabel
    }%
}
\makeatother

\theoremstyle{definition}

\crefname{claim}{claim}{claims}
\crefname{inftheorem}{informal theorem}{informal theorems}
\crefname{figure}{figure}{figures}
\crefname{@theorem}{theorem}{theorems}
\crefname{Definition}{definition}{definitions}
\crefname{line}{line}{lines}

\Crefname{claim}{Claim}{Claims}
\Crefname{inftheorem}{Informal Theorem}{Informal Theorems}
\Crefname{@theorem}{Theorem}{Theorems}
\Crefname{Definition}{Definition}{Definitions}
\Crefname{figure}{Figure}{Figures}
\Crefname{line}{Line}{Lines}

\newcommand{\R}{\mathbb{R}}

\newcommand{\E}{\mathbb{E}}
\newcommand{\D}{\mathcal{D}}
\newcommand{\eps}{\epsilon}

\newcommand{\Alg}{\mathsf{Alg}}
\newcommand{\Off}{\mathsf{Off}}
\newcommand{\SSAlg}{\mathsf{SSAlg}}
\newcommand{\Opt}{\mathsf{Opt}}

\newcommand{\supp}{\operatorname*{supp}}

\newcommand{\RSG}{\mathsf{RSG}}

\newcommand{\va}{\bm{a}}
\newcommand{\vc}{\bm{c}}
\newcommand{\vp}{\bm{p}}
\newcommand{\vq}{\bm{q}}
\newcommand{\vr}{\bm{r}}
\newcommand{\vs}{\bm{s}}
\newcommand{\vT}{\bm{T}}

\newcommand{\vv}{\bm{v}}
\newcommand{\vx}{\bm{x}}
\newcommand{\vy}{\bm{y}}
\newcommand{\vz}{\bm{z}}

\allowdisplaybreaks

\algrenewcommand\algorithmiccomment[1]{\hfill {\color{gray} \(\triangleright\) \itshape #1}}
\algnewcommand{\LineComment}[1]{\item[] {\color{gray}  \(\triangleright\) \itshape #1}}
\algnewcommand\Input{\item[\textbf{Input:}]}
\algnewcommand\Output{\item[\textbf{Output:}]}
\algnewcommand{\IfThenElse}[3]{
  \State \algorithmicif\ #1\ \algorithmicthen\ #2\ \algorithmicelse\ #3}

\algnewcommand{\IfThen}[2]{
\State \algorithmicif\ #1\ \algorithmicthen\ #2}

\newcommand{\remove}[1]{}

\AddToHook{cmd/appendix/before}{%
    \crefalias{section}{appendix}%
    \crefalias{subsection}{appendix}
}

\title{A Constant-Factor Single-Sample Prophet Inequality\\ for Subadditive Valuations}

\author{
Shuchi Chawla \\ \small The University of Texas at Austin \\ {\tt \small shuchi@cs.utexas.edu} \and 
Trung Dang  \\ \small The University of Texas at Austin \\ {\tt \small dddtrung@cs.utexas.edu}
}

\date{}

\begin{document}

\maketitle
\thispagestyle{empty}

\begin{abstract}
We establish a constant-factor single-sample prophet inequality for online combinatorial allocation with monotone subadditive valuations. Given one independent sample from each buyer's unknown distribution, our reduction converts any offline algorithm achieving an $\alpha$ fraction of optimal welfare into an online algorithm achieving an $\alpha/120$ fraction of the expected offline optimum. The guarantee holds for arbitrary arrival orders and extends to the stronger Game of Googol model. Our approach has two key ingredients: a {\em  prior-free} random score generator and a new sample-based construction of tentative allocations. The tentative allocations capture substantial welfare while controlling contention for items, and the score generator relates itemwise competition in the tentative allocations to the value of retained bundles. The reduction is efficient whenever its offline allocator and score generator are efficient.
\end{abstract}

\newpage
\addtocounter{page}{-1}
\section{Introduction}
\label{sec:intro}

In a combinatorial auction, the goal is to allocate a set of $m$ items among $n$ buyers with combinatorial valuations over bundles of items to maximize social welfare: the sum of the buyers' values for their assigned bundles. In the online setting, buyers arrive sequentially. Buyer $i$'s valuation $v_i:2^{[m]}\to\R_{\geq0}$ is drawn independently from a buyer-specific distribution $\mathcal D_i$. When she arrives, her valuation is revealed, and the algorithm must irrevocably allocate to her a subset of the remaining items. A \emph{prophet inequality} compares the algorithm's expected welfare with the expected welfare of an optimal allocation chosen with full knowledge of all realized valuations. This hindsight optimum serves as the \emph{prophet} benchmark.

A central question is how much information about the underlying distributions $\mathcal D_i$ is needed to compete with the prophet. Single-sample prophet inequalities push this question to its informational extreme: the distributions are unknown, and the algorithm receives only one independent sample valuation from each buyer's distribution before arrivals begin. Remarkably, for a single item, one sample suffices for the optimal $1/2$ welfare guarantee~\cite{rubinstein20samples}. Constant-factor single-sample guarantees are also known for several combinatorial settings. Notably, D\"utting et al.~\cite{dutting2024online} established such guarantees for combinatorial allocation with XOS (fractionally subadditive) valuations. Our recent work~\cite{chawla2026single} gives a modular reduction from this setting to single-item prophet inequalities, using buyer-local tentative allocations, supporting prices, and a free-disposal benchmark.

In this paper, we study \emph{subadditive} valuations, which satisfy $v(S\cup T)\leq v(S)+v(T)$ for $S, T\subseteq [m]$, and form a strict superclass of XOS valuations. For general subadditive valuations, constant-factor prophet inequalities have required substantially more distributional information. Following the $O(\log\log m)$ guarantee of D\"utting, Kesselheim, and Lucier~\cite{dutting2020prophet}, Correa and Cristi~\cite{correa2023constant} proved the existence of a constant-factor guarantee. More recently, MacRury et al.~\cite{macrury2026efficient} and Kroll et al.~\cite{KMRS26}  obtained efficient constant-factor prophet inequalities using demand queries and sampling access to the distributions, with polynomially many samples in expectation. Whether a single sample suffices for a constant-factor guarantee for general subadditive valuations remained open. {\bf We resolve this question affirmatively.} 

\begin{inftheorem} \label{thm:informal}
There exists an online allocation algorithm $\SSAlg$ using one sample per buyer such that
\[
\E[\SSAlg]\geq\frac{1}{120}\E[\Opt].
\]
The guarantee holds for monotone subadditive valuations and arbitrary buyer arrival orders, including orders chosen after the valuations are realized.
\end{inftheorem}

The guarantee extends to the stronger Game of Googol model, where an adversary specifies two valuations per buyer and an independent fair coin determines which is the sample and which is the real valuation. Our reduction also accommodates approximate offline allocation. Given an offline algorithm that achieves an $\alpha$ fraction of the optimal welfare, our single-sample prophet inequality achieves an $\alpha/120$ fraction of the prophet benchmark. In particular, Feige's efficient $2$-approximation \cite{feige2009maximizing} yields a $240$-competitive guarantee.

Our single-sample algorithm is not necessarily computationally efficient. Even with an efficient offline allocator, the running time depends on a second ingredient, the prior-free random score generator (RSG) described below. We can construct an RSG in exponential time by solving a linear program whose variables specify a distribution over discretized score vectors. We show that constructing an RSG with a constant-factor guarantee requires, in general, exponentially many value and demand queries to the underlying valuation function. On the positive side, we identify a class of subadditive valuations that admit efficient score generation. The resulting polynomial-time single-sample algorithm achieves a constant competitive ratio even for valuations in this class whose best XOS approximation loses a factor of $\Theta(\log m)$.

\subsection{Overview of the Approach}

Our approach has two key ingredients. The first is a random score generator (RSG) that, given a valuation $v$ and a tentative bundle $X$, assigns a score to each item in $X$. This generator is inspired by Correa and Cristi's subadditive prophet inequality~\cite{correa2023constant}, but is prior-free: it depends only on $v$ and $X$, and not on the value distributions. The second ingredient is a
sample-based construction of tentative allocations that simultaneously capture substantial value and control competition for individual items. This construction builds on the framework developed in our single-sample prophet inequality for XOS valuations~\cite{chawla2026single}. We describe these ingredients before explaining how to implement the resulting allocation online.

\paragraph{A distribution-independent score generator.} The purpose of an RSG is to relate competition for individual items to the value of the bundle that a buyer ultimately receives. In a hypothetical allocation, each buyer wins the items on which her score exceeds all competing scores, much as bids determine the allocation in a simultaneous first-price auction. The challenge is to choose scores so that resolving competition item by item preserves a substantial fraction of the tentative welfare.

Our construction builds on the bidding argument of Feldman et al.~\cite{FFGL13}. Given a subadditive valuation $v$, a bundle $X$, and any $\varepsilon>0$, our RSG returns a nonnegative vector $\vq$, supported on $X$, such that
\begin{equation}
\E_{\vq}\!\left[v\bigl(\{j\in X:q_j>p_j\}\bigr)\right]
\geq\left(\frac12-\epsilon\right) v(X)+\E_{\vq}[\vq(X)]-\vp(X)
\qquad\text{for every item pricing }\vp\geq0.
\label{eq:intro-score}
\end{equation}

Here $\vq(S)=\sum_{j\in S}q_j$. Its existence follows from a simple symmetry argument along with von Neumann's minimax theorem: the symmetric bidding argument gives a suitable response to each distribution of competing prices, and minimax yields a single score distribution that works against every price vector \cite{FFGL13}.

To see how this guarantee helps, let $X_i$ be buyer $i$'s tentative bundle, and independently generate scores $\vq_i$ from the RSG. Let $\vp_i$ be the coordinate-wise maximum of the other buyers' scores, and let $W_i=\{j\in X_i:q_{i,j}>p_{i,j}\}$ be buyer $i$'s hypothetical winning set. Applying \eqref{eq:intro-score} to each buyer and summing gives a lower bound on the expected welfare of the winning sets: a constant fraction of the tentative welfare, plus the total scores, minus the competing prices charged to the tentative bundles.

The last two terms admit a useful accounting interpretation. Each score is credited to the buyer who generates it, but may also contribute to the prices faced by other buyers whose tentative bundles contain that item. We therefore seek a bound on the total competing-price charge relative to the total scores. Matching this bound with the coefficient of the positive score term makes the two contributions cancel. The required notion of contention is thus weighted by the scores and bounded in expectation.

The tentative allocations $X_i$ must consequently satisfy two properties: they must capture a constant fraction of the prophet's welfare, and they must have sufficiently controlled contention. To see how these properties can arise, let $\Opt_i(v_i)$ denote the distribution over bundles assigned to buyer $i$ in an optimal offline allocation when her valuation is $v_i$ and all other valuations are drawn independently from their distributions. If each buyer independently draws $X_i$ from $\Opt_i(v_i)$, the expected tentative welfare equals the prophet's welfare, and each item belongs to at most one tentative bundle in expectation. Independence then gives the required bound on competing-price charges. The same reasoning applies to an approximate offline allocator, with the corresponding loss in tentative welfare. Our task is to reproduce these properties using only a few samples.

\paragraph{Replacing distributional information by two-entry banks.} For simplicity, suppose we have three independent samples per buyer. We use one as a \emph{competition sample} to set the online thresholds, and the other two as \emph{coordination samples} to generate tentative allocations. The competition sample allows the thresholds to reflect competition from buyers who have not yet arrived.

Give each buyer a bank $B_i=\{c_i^0,c_i^1\}$ containing her two coordination samples. To generate buyer $i$'s tentative allocation, we form a synthetic instance: insert $v_i$ at coordinate $i$, and independently select a uniform entry of $B_k$ at each coordinate $k\ne i$. We run the offline allocator on this instance and retain buyer $i$'s bundle. Averaging over the banks, each synthetic instance has the correct distribution, so the tentative welfare guarantee is preserved.

Reusing the banks, however, correlates the tentative allocations of different buyers. In particular, a large score on an item may coincide with many competing requests for that item. The two bank entries let us control these correlations by bounding the score-weighted demand for each item. This lets us bound the expected excess of the competing-price charges over the total scores by a constant times the expected sum of the winning scores. Combining with the score generator, we show that the winning sets retain a constant fraction of the tentative welfare in expectation.

\paragraph{From hindsight to online allocation.} We apply the same tentative-allocation and scoring procedure to each competition-sample valuation, using fresh randomness, and set each item's threshold to the largest resulting score. When a real buyer arrives, we generate her tentative bundle and scores, and allocate every still-available item in that bundle whose score exceeds its threshold.

To analyze this rule, we adapt the sample-symmetry argument of~\cite{rubinstein20samples,chawla2026single}. Conditional on the banks, the sample and real evaluations of each buyer are exchangeable. Competition against the sample yields two identically distributed subsets of each buyer's winning set that together cover it; one of these subsets is guaranteed to be allocated to her online. Conditional on a buyer's evaluation being real, subadditivity therefore guarantees that she receives at least half the value of her winning set in expectation.


Together, these ingredients give a constant-factor prophet inequality using three samples per buyer. The guarantee also holds in the stronger Game of Googol model, allowing us to apply the sample-reduction theorem of D\"utting et al.~\cite[Theorem~3.10]{dutting2024online}. This yields a single-sample prophet inequality at a loss of a factor of $4$ in the competitive ratio.




\subsection{Related Work}
\paragraph{Combinatorial prophet inequalities.}
Prophet inequalities have been established for a broad range of
combinatorial constraints and objectives. Kleinberg and
Weinberg~\cite{KW12} obtained the optimal factor-two guarantee for
linear objectives under matroid constraints, while Rubinstein and
Singla~\cite{RS17} obtained constant-factor guarantees for submodular
objectives under matroid constraints and polylogarithmic guarantees
for subadditive objectives under general downward-closed constraints.
For online combinatorial allocation, Feldman, Gravin, and
Lucier~\cite{FGL15} established the optimal factor-two guarantee for
XOS valuations using static anonymous item prices. D\"utting
et al.~\cite{DFKL20} developed pricing guarantees for more general
combinatorial markets, multidimensional matroids, and packing problems.
For subadditive buyers, Zhang~\cite{Zha20} obtained an
$O(\log m/\log\log m)$ guarantee, and D\"utting, Kesselheim, and
Lucier~\cite{dutting2020prophet} obtained an $O(\log\log m)$ guarantee.
Correa and Cristi~\cite{correa2023constant} subsequently established
the first constant-factor guarantee, with competitive ratio $6$.
More recently, MacRury et al.~\cite{macrury2026efficient} gave an
expected-polynomial-time $(6+\varepsilon)$-competitive algorithm using
demand queries, Kroll et al.~\cite{KMRS26} obtained an efficient
$4$-competitive algorithm for subadditive combinatorial allocation, and Cristi and Zhang~\cite{cristi2026optimal} showed that online subadditive combinatorial allocation does not admit a $2$-competitive algorithm.

\paragraph{Single-sample and few-sample guarantees.}
Azar, Kleinberg, and Weinberg~\cite{AKW14} initiated the study of
prophet inequalities with limited distributional information.
Rubinstein, Wang, and Weinberg~\cite{rubinstein20samples} showed that
one sample suffices for the optimal single-item guarantee, while
Correa et al.~\cite{CCES20} studied the related two-sided Game of
Googol model. Constant-factor single-sample guarantees are known for
matching, several classes of matroids, and budget-additive
combinatorial allocation~\cite{AKW14,KNR22,CDF22}. For general
matroids, Fu et al.~\cite{FuEtAl24} obtained a constant-factor
guarantee using polylogarithmically many samples. D\"utting
et al.~\cite{dutting2024online} established the first constant-factor
single-sample guarantee for general XOS valuations, as well as a
truthful $(2+\varepsilon)$-competitive mechanism using polynomially
many samples. Chawla and Dang~\cite{chawla2026single} improved the
single-sample competitive ratio for XOS valuations to $10.4$ and also
obtained multi-sample guarantees and extensions to divisible resources
with capped-XOS valuations. For XOS buyers in a fixed known arrival
order, Rubinstein and Singla~\cite{RS26} obtained a truthful
$(2+\varepsilon)$-approximation using $O(\log\log m)$ samples for
constant $\varepsilon>0$, and a truthful $O(\log\log m)$-approximation
using one sample. For general subadditive valuations, the efficient
constant-factor algorithm of MacRury et al.~\cite{macrury2026efficient}
uses polynomially many samples in expectation. In the large-capacity
regime, Ghuge, Singla, and Wang~\cite{GSW25} and Faw, Singla, and
Wang~\cite{FSW26} obtained single-sample resource-allocation algorithms
whose welfare approaches the expected offline optimum as resource
capacities grow.

\subsection*{AI Disclosure}
\addcontentsline{toc}{subsection}{AI Disclosure}
GPT-6.2 Astra and Gemini 4 Neon were used to help develop the results in this paper and polish their presentation. In particular, Astra proposed the score generator of \Cref{lem:score-generator} after being asked to approach the problem of online combinatorial allocation with reallocation. This was then used to lift any online reallocation algorithm (and eventually any offline algorithm) to the constant-sample setting. \Cref{thm:three-sample} is a simplified and streamlined version of this argument, and we did not try to optimize the constant ratio for the sake of exposition. Astra also proved the computational hardness result and constructed the subadditive subclass presented in \Cref{sec:score-efficiency}. The authors then simplified the arguments for both results. The authors take full responsibility for the correctness and presentation of this paper.

\section{Preliminaries}
\label{sec:prelims}

\paragraph{Online combinatorial allocation.} There are $n$ buyers indexed by $[n]$ and $m$ indivisible items indexed by $[m]$. Buyer $i$ has a valuation $v_i:2^{[m]}\to\R_{\ge 0}$, and we write $\vv=(v_1,\ldots,v_n)$ for a valuation profile. An allocation $(A_1,\ldots,A_n)$ is feasible if the bundles $A_i\subseteq[m]$ are pairwise disjoint. Its welfare on $\vv$ is $\sum_{i=1}^n v_i(A_i)$. We write $\Opt(\vv)$ for the maximum welfare over feasible allocations and $\Off_i(\vv)$ for the bundle assigned to buyer $i$ by an offline allocation algorithm $\Off$.

Buyers arrive sequentially. When buyer $i$ arrives, the algorithm observes the buyer's real valuation $r_i$ and irrevocably allocates a bundle $A_i$ from the remaining items. For the real profile $\vr=(r_1,\ldots,r_n)$, we compare the algorithm's expected welfare $\E[\Alg]$, where $\Alg=\sum_{i=1}^n r_i(A_i)$, with the expected hindsight optimum $\E[\Opt(\vr)]$. The expectations are over the sample and real profiles and the algorithm's randomness.

\paragraph{Sample-based setting.} Each real valuation $r_i$ is drawn independently from an unknown buyer-specific distribution $\D_i$. In the $k$-sample setting, the algorithm receives $k$ sample profiles $\vs^{(1)},\ldots,\vs^{(k)}$ before the buyers arrive. These profiles and $\vr$ are independent draws from $\D=\prod_{i=1}^n\D_i$. We write $\vs=(s_1,\ldots,s_n)$ for the sample profile when $k=1$.

\paragraph{Game of Googol.} In the $k$-sample Game of Googol, an adversary fixes $k+1$ indexed valuations for each buyer. Distinct indices may represent the same valuation function. Independently for each buyer, a uniform random permutation assigns the indices to $k$ sample slots and one real slot. The algorithm receives the sample profiles before any arrivals and observes $r_i$ when buyer $i$ arrives. Throughout the rest of the paper, we adopt the Game of Googol setting whenever we talk about the sample-based setting.

\paragraph{Valuations.} Let $\mathcal V$ denote the class of normalized, monotone, subadditive valuations on $[m]$. Thus $v\in\mathcal V$ satisfies $v(\varnothing)=0$, $v(S)\le v(T)$ whenever $S\subseteq T\subseteq[m]$, and $v(S\cup T)\le v(S)+v(T)$ for all $S,T\subseteq[m]$. Throughout, we take all sample and real valuations to belong to $\mathcal V$.

\paragraph{Value and demand queries.} For a valuation $v$ and an item set $X\subseteq[m]$, a value query returns $v(S)$ for a set $S\subseteq X$, and a demand query at prices $\vp$ returns a bundle maximizing $v(S)-\vp(S)$ over $S\subseteq X$.

\paragraph{Notation.} For an item vector $\vq=(q_j)_{j\in[m]}$ and a bundle $S\subseteq[m]$, let $\vq(S)=\sum_{j\in S}q_j$. We use this notation for any item vector and any subset of its index set. A vector indexed by $[m]$ is supported on $X\subseteq[m]$ if its coordinates outside $X$ are zero. We write $[k]$ for the set $\{1, \dots, k\}$, $\Delta(B)$ for the set of probability distributions on $B$, $\supp(\mu)$ for the set of outcomes of positive probability when $\mu$ is a distribution on a finite set, and $\mathbf{1}[E]$ for the indicator of an event $E$.

\section{A Prior-Free Score Generator}
\label{sec:prior-free-scores}

Our first ingredient is a score generator for subadditive valuations, following previous approaches for subadditive online combinatorial allocation~\cite{correa2023constant,macrury2026efficient,KMRS26}. Our scores are \emph{prior-free} and depend only on the valuation and the tentative bundle. This allows such a score generator to be used in the single-sample setting.

\begin{lemma}
\label{lem:score-generator}
For any $\eps \in (0, 1/2)$, there is a random score generator $\varphi_\eps : \mathcal{V} \times 2^{[m]} \to \Delta(\R_{\ge 0}^m)$ such that the following is true. Let $v \in \mathcal{V}$ be a normalized, monotone, subadditive valuation and $X\subseteq [m]$ be a tentative allocation set. Draw the random score vector $\vq \sim \varphi_\eps(v, X)$. Then, for every $\vp\in\R_{\ge 0}^m$,
\[
\E\!\left[v\bigl(\{j\in X:q_j>p_j\}\bigr)\right]
\ge \left(\frac12-\eps\right)v(X)+\E[\vq(X)]-\vp(X).
\]
Moreover, with probability one, $\vq(S)\le v(S)+\eps v(X)$ simultaneously for all $S\subseteq X$.
\end{lemma}


\begin{proof}
Let $V = v(X)$ be the bound on $v$. If $V=0$, the zero vector satisfies both inequalities. Otherwise, set $\delta=\eps V/|X|$ and let $\mathcal G = \left\{0, \delta, 2\delta, \dots, \lceil \frac{V}{\delta} \rceil \cdot \delta\right\}^X$ be the finite set of nonnegative vectors on $X$ whose coordinates lie on a $\delta$-grid and are bounded by $V$ up to rounding.

Consider a zero-sum game in which we choose a distribution $\mu\in\Delta(\mathcal G)$ over score vectors and the opponent chooses a competing distribution $\nu\in\Delta(\mathcal G)$. Our payoff is
\[
U(\mu,\nu)=\E_{\vx \sim \mu, \vy \sim \nu}[v\bigl(\{j\in X:x_j\ge y_j\}\bigr)-\vx(X)+\vy(X)].
\]

For now, we break ties between $x_j$ and $y_j$ in our favor. Observe that $\{j\in X:x_j\ge y_j\}$ and $\{j\in X:y_j\ge x_j\}$ together cover $X$. By subadditivity, $v\left(\{j\in X:x_j\ge y_j\}\right) + v\left(\{j\in X:y_j\ge x_j\}\right) \ge V$, and therefore $U(\mu, \nu) + U(\nu, \mu) \ge V$. This shows that
\[
\min_{\nu \in \Delta(\mathcal{G})} \max_{\mu\in \Delta(\mathcal{G})} U(\mu,\nu)\ge V/2
\]
as we can choose $\mu = \nu$ if the opponent goes first and guarantee a payoff of at least $V/2$. As $\mathcal{G}$ is finite, we then apply von Neumann's minimax theorem to obtain
\[
\max_{\mu\in \Delta(\mathcal{G})} \min_{\nu \in \Delta(\mathcal{G})} U(\mu,\nu)\ge V/2,
\]
so the maximin choice $\mu^*$ gives a payoff of at least $V/2$ against any pure score vector $\vy\in\mathcal G$. The guarantee extends to every nonnegative $\delta$-grid vector $\vy$ on $X$. If $\vy\notin\mathcal G$, then one coordinate exceeds $V$, so $U(\mu^*,\vy)\ge \E_{\vx \sim \mu^*}[\vy(X)-\vx(X)]>U(\mu^*,\mathbf{0})$.

Let $\vx \sim \mu^*$ and set $q_j=x_j+\delta$ on $X$, with zero scores elsewhere. For any competing vector $\vp$, round each coordinate on $X$ down to a multiple of $\delta$ to get $\bar{\vp}$. Every weak win against $\bar{\vp}$ is now a strict win against $\vp$, since $x_j\ge \bar p_j$ implies $q_j\ge \bar p_j+\delta>p_j$. The shift adds $|X|\delta$ to our score total, and rounding can only decrease the competing total. By monotonicity,
\[
\E\!\left[v\bigl(\{j\in X:q_j>p_j\}\bigr)\right]-\E[\vq(X)]+\vp(X)\ge U(\mu^*,\bar{\vp})-|X|\delta\\
\ge \left(\frac12-\eps\right)V.
\]

For the subset bound, suppose a vector $\vx\in\supp(\mu^*)$ satisfies $\vx(S)>v(S)$ for some $S\subseteq X$, and let $\vx'$ be the grid vector obtained by setting its scores on $S$ to zero. Fix a competing vector $\vy$ and let $W$ and $W'$ be the bundles won by $\vx$ and $\vx'$, respectively. The scores agree outside $S$, so $W\subseteq W'\cup S$. Monotonicity and subadditivity give $v(W)\le v(W'\cup S)\le v(W')+v(S)$. The score total decreases by $\vx(S)$, so replacing $\vx$ by $\vx'$ improves the payoff by at least $\vx(S)-v(S)>0$ against every $\vy$. Therefore, transferring probability mass from $\vx$ to $\vx'$ increases the minimum expected payoff against $\vy \in \mathcal{G}$, contradicting the maximin choice of $\mu^*$. Hence every draw from $\mu^*$ satisfies $\vx(S)\le v(S)$ simultaneously for all $S\subseteq X$. This proves the claim, since $\vq(S)=\vx(S)+|S|\delta\le v(S)+\eps V$ for each such $S$.
\end{proof}

\section{A Constant-Sample Combinatorial Prophet Inequality}
\label{sec:three-sample}

We now turn to designing an online combinatorial allocation algorithm by lifting \emph{any} offline allocation algorithm to the online setting using only \emph{three samples}. We follow an approach similar to that of \cite{chawla2026single} for two-sample combinatorial allocation with XOS buyers. In particular, the set of samples contains two \emph{coordination samples} $\vc^0=(c_1^0,\dots,c_n^0)$ and $\vc^1=(c_1^1,\dots,c_n^1)$ and a \emph{competition sample} $\vs$. We write $\vr=(r_1,\dots,r_n)$ for the real valuations. Furthermore, suppose we have access to an offline allocation algorithm $\Off$ that takes any full instance $\vv = (v_1, \dots, v_n)$ and returns a bundle $\Off_i(\vv)$ for each buyer $i$. We assume for now that $\Off$ is deterministic.

\begin{enumerate}
  \item First, imagine running $\Off$ on a random instance formed from the coordination samples by choosing each buyer $i$'s valuation uniformly from $c_i^0$ and $c_i^1$. Then, for each sample buyer $i$, we imagine replacing the $i$-th valuation in this run with $s_i$. This gives sample buyer $i$ a tentative sample bundle $X_i$. We do the same for real buyers to obtain a tentative real bundle $X_i'$.
  \item We then draw a sample score vector $\vq_i \sim \varphi_\eps(s_i, X_i)$ and a real score vector $\vq'_i \sim \varphi_\eps(r_i, X_i')$, using the score generator $\varphi_\eps$ from \Cref{lem:score-generator} and some parameter $\eps$ to be chosen later.
  \item Finally, we leverage the threshold algorithm of \cite{rubinstein20samples} for the single-item, single-sample prophet inequality. We take $p_{j} = \max_{i \in [n]} q_{i, j}$ to be the maximum score for item $j$ and then allocate item $j$ to the first real buyer whose score $q'_{i, j}$ crosses $p_{j}$.
\end{enumerate}

\Cref{alg:three-sample} gives the formal algorithm. For simplicity, we assume that each item's positive scores are distinct. Standard techniques handle ties.

\begin{algorithm}[hbt!]
\caption{Three-Sample Combinatorial Online Allocation}
\label{alg:three-sample}
\begin{algorithmic}[1]
\Input{Sample valuation vectors $(\vc^0, \vc^1, \vs)$ and arriving buyers with valuations $\vr$.}
\Output{A feasible allocation $A_1, \dots, A_n$.}
\State Use $\vc^0$ and $\vc^1$ as the coordination samples for \textsc{Evaluate} (\Cref{alg:three-sample-evaluate}).
\For{each buyer $i$}
    \State $(X_i,\vq_i)\gets\Call{Evaluate}{i,s_i}$.
\EndFor
\State Let $p_j \gets \max_{k \in [n]} q_{k, j}$ for all $j \in [m]$.
\State Let $R \gets [m]$ be the set of available items.
\For{each real buyer $i$ upon arrival}
    \State $(X_i',\vq'_i)\gets\Call{Evaluate}{i,r_i}$.
    \State Allocate the bundle $A_i\gets\{j\in R:q'_{i,j}>p_j\}$ to buyer $i$.
    \State $R\gets R\setminus A_i$.
\EndFor
\end{algorithmic}
\end{algorithm}

\begin{algorithm}[hbt!]
\caption{\textsc{Evaluate}$(i,v)$}
\label{alg:three-sample-evaluate}
\begin{algorithmic}[1]
\Input{Buyer $i$ with valuation $v$.}
\Output{A tentative bundle $X$ and its random score vector $\vq$.}
\For{each buyer $k\ne i$}
    \State $z_k \gets c_k^0$ or $z_k \gets c_k^1$ uniformly at random and independently of the choices for other buyers.
\EndFor
\State $z_i \gets v$.
\State $X\gets \Off_i(\vz)$, where $\vz=(z_1,\ldots,z_n)$ is called a \emph{synthetic instance}.
\State Draw $\vq\sim\varphi_\eps(v,X)$ with fresh randomness.
\State \Return $(X,\vq)$.
\end{algorithmic}
\end{algorithm}

We first prove the following theorem.



\begin{theorem}
\label{thm:three-sample}
Let $\Off$ be any deterministic feasible offline allocation algorithm for subadditive valuations. In the three-sample Game of Googol, for a suitable choice of $\eps$, \Cref{alg:three-sample} satisfies
\[
\E[\Alg]\ge \frac1{30} \cdot \E\!\left[\sum_{i=1}^n r_i\bigl(\Off_i(\vr)\bigr)\right].
\]
\end{theorem}

In \Cref{subsec:three-to-one-sample}, we extend this result to obtain a single-sample online combinatorial allocation algorithm, given access to any (randomized) offline allocation algorithm. In particular, we obtain the following corollary.

\begin{corollary}[\Cref{thm:informal} restated]\label{corr:ssalg}
  Let $\Off$ be any (random) feasible offline allocation algorithm for subadditive valuations that is an $\alpha$-approximation to the hindsight optimum $\Opt$ on any subadditive instance. Then there exists a single-sample Game-of-Googol online combinatorial allocation algorithm $\SSAlg$ for subadditive valuations such that
  \[
  \E[\SSAlg]\ge \frac{\alpha}{120} \cdot \E_{\vr}[\Opt(\vr)].
  \]

  In particular, using the offline optimal allocation algorithm as $\Off$, we obtain a $120$-competitive single-sample online combinatorial allocation algorithm.
\end{corollary}


\subsection{Welfare and Competition of Tentative Allocations} \label{subsec:tentative}

We first look at the \emph{total tentative welfare} $T$ across all tentative allocations $X_i$ and $X'_i$, defined by
\[T = \sum_{i=1}^n \bigl(s_i(X_i)+r_i(X'_i)\bigr).\]

Let $\vz^{(i)}$ and $\vz^{(i)\prime}$ be the synthetic instances used by \Cref{alg:three-sample-evaluate} to compute $X_i$ and $X'_i$. It is straightforward to see that $\vz^{(i)}$ and $\vz^{(i)\prime}$ are distributed identically to $\vr$, so $\E[s_i(X_i)]=\E[r_i(X'_i)]=\E[r_i(\Off_i(\vr))]$. Summing over buyers gives $\E[T]=2\E\!\left[\sum_{i=1}^n r_i\bigl(\Off_i(\vr)\bigr)\right]$.

For every item $j$, let $d_j=\sum_{i=1}^n(\mathbf{1}[j\in X_i]+\mathbf{1}[j\in X_i'])$ be the total tentative demand for this item. The following lemma shows that for any buyer $i$ who tentatively requests item $j$, the \emph{weighted competition} on this item is not too far from their own score.


\begin{lemma}
\label{lem:three-sample-competition}
For every buyer $i$ and item $j$, $\E[d_jq_{i,j}]\le 5\E[q_{i,j}]$ and $\E[d_jq'_{i,j}]\le 5\E[q'_{i,j}]$.
\end{lemma}

\begin{proof}
We focus on proving the first statement, as the second statement follows by symmetry. First sample two extra instances $\vy, \vy'$ as follows.

\begin{itemize}
  \item For each buyer $i$, independently choose two valuations uniformly at random from the pool of four fixed valuations $v_{i, 1}, \dots, v_{i, 4}$ to form the \emph{faux} coordination samples $\vc^{0\prime}$ and $\vc^{1\prime}$.
  \item Then, for each buyer $i$, draw $y_i$ and $y'_i$ independently, choosing either $c_i^{0\prime}$ or $c_i^{1\prime}$ with probability $1/2$ each.
\end{itemize}

We then let $E_k = \{y_k \neq y'_k\}$ be the event that the two instances use different valuations for buyer $k$. The events $E_k$ are independent and each has probability $1/2$. Finally, draw $\widetilde{\vq}_i \sim \varphi_\eps(y_i,\Off_i(\vy))$ with fresh randomness. Observe that, conditional on $E_i$, $y_i$ is still drawn uniformly from $c_i^{0\prime}$ and $c_i^{1\prime}$, so $\vy$ is distributed identically to $\vz^{(i)}$ and $\E[\widetilde q_{i,j}\mid E_i]=\E[q_{i,j}]$.

We now fix any other buyer $k \neq i$. At coordinate $i$, $z^{(i)}_i$ is the competition sample and $z^{(k)}_i$ is a coordination sample, so $z^{(i)}_i \neq z^{(k)}_i$. At coordinate $k$, $z^{(i)}_k$ is a coordination sample and $z^{(k)}_k$ is the competition sample, so we also have $z^{(i)}_k \neq z^{(k)}_k$. Finally, for all other coordinates $\ell \notin \{i, k\}$, $z^{(i)}_\ell$ and $z^{(k)}_\ell$ are independent draws from the coordination samples. This implies that the pair $(\vz^{(i)},\vz^{(k)})$ is distributed identically to the pair $\bigl((\vy,\vy')\mid E_i\cap E_k\bigr)$. The same argument shows that the pair $(\vz^{(i)},\vz^{(k)\prime})$ is distributed identically to the pair $\bigl((\vy,\vy')\mid E_i\cap E_k\bigr)$. The weighted competition that buyer $k$ imposes on buyer $i$ is then
\begin{align*}
  \E\!\left[q_{i,j}\bigl(\mathbf{1}[j\in X_k]+\mathbf{1}[j\in X'_k]\bigr)\right] &= \E\!\left[q_{i,j}\mathbf{1}[j\in X_k]\right] + \E\!\left[q_{i,j}\mathbf{1}[j\in X'_k]\right] \\
  &= 2 \E\!\left[\widetilde q_{i,j}\mathbf{1}[j\in \Off_k(\vy')] \mid E_i \cap E_k\right] \\ 
  &=4\E\!\left[\widetilde q_{i,j}\mathbf{1}[E_k]\mathbf{1}[j\in\Off_k(\vy')]\mid E_i\right],
\end{align*}
where the last equality uses $\Pr[E_k \mid E_i] = 1/2$.

We now consider buyer $i$'s own tentative bundles. At coordinate $i$, $z^{(i)}_i$ is the competition sample and $z^{(i)\prime}_i$ is the real valuation, so $z^{(i)}_i \neq z^{(i)\prime}_i$. For every other coordinate $\ell\ne i$, $z^{(i)}_\ell$ and $z^{(i)\prime}_\ell$ are independent draws from the coordination samples. This implies that the pair $(\vz^{(i)},\vz^{(i)\prime})$ is distributed identically to the pair $\bigl((\vy,\vy')\mid E_i\bigr)$. Then the weighted competition from buyer $i$ is
\begin{align*}
  \E\!\left[q_{i,j}\bigl(\mathbf{1}[j\in X_i]+\mathbf{1}[j\in X'_i]\bigr)\right] &= \E\!\left[q_{i,j}\mathbf{1}[j\in X_i]\right] + \E\!\left[q_{i,j}\mathbf{1}[j\in X'_i]\right] \\
  &= \E[q_{i,j}] + \E\!\left[\widetilde q_{i,j}\mathbf{1}[j\in \Off_i(\vy')]\mid E_i\right].
\end{align*}

Summing over all buyers gives
\begin{align*}
  \E[d_jq_{i,j}] &= \E[q_{i,j}] + \E\!\left[\widetilde q_{i,j}\mathbf{1}[j\in \Off_i(\vy')]\mid E_i\right] + 4\sum_{k\ne i}\E\!\left[\widetilde q_{i,j}\mathbf{1}[E_k]\mathbf{1}[j\in \Off_k(\vy')]\mid E_i\right] \\
  &\le \E[q_{i,j}] + 4\E\!\left[\widetilde q_{i,j}\sum_{k=1}^n\mathbf{1}[j\in \Off_k(\vy')]\mid E_i\right] \\
  &\le \E[q_{i,j}] + 4\E[\widetilde q_{i,j}\mid E_i] = 5\E[q_{i,j}],
\end{align*}
where the second inequality holds because $\Off(\vy')$ assigns item $j$ to at most one buyer, and the final equality follows from $\E[\widetilde q_{i,j}\mid E_i]=\E[q_{i,j}]$.
\end{proof}

\subsection{Welfare of the Highest-Scoring Proposers}
\label{subsec:three-sample-highest}

We now consider the welfare obtained from the highest-scoring proposers for each item. In particular, let $\mathcal{A}$ be the set of $2n$ tentative tuples consisting of $(s_i, X_i, \vq_i)$ and $(r_i, X'_i, \vq'_i)$ for each buyer $i$. Denote any tuple in $\mathcal{A}$ by $(u_a, X_a, \vq_a)$. For each $a\in\mathcal A$, let $W_a$ consist of the items in $X_a$ on which $a$ has the highest score, so $W_a=\{j\in X_a:q_{a,j}>\max_{b\in\mathcal A\setminus\{a\}}q_{b,j}\}$.

Let $H = \sum_{a \in \mathcal A} u_a(W_a)$ be the welfare of the highest-scoring proposers. The following lemma shows that we incur only a constant-factor loss in welfare.

\begin{lemma} \label{lem:welfare-highest-score}
  $\E[H] \ge \left(\frac{2}{29} - \eps\right) \E[T]$, where $\eps$ is the precision parameter of \Cref{alg:three-sample}. In particular, for $\eps = 1/435$, we have $\E[H] \ge \frac{\E[T]}{15}$.
\end{lemma}

\begin{proof}
  For each tuple $a \in \mathcal{A}$ and item $j$, define the rival score on this item as $\widehat p_{a,j}=\max(\{q_{b,j}:b\in\mathcal A\setminus\{a\}\})$. By definition, we have $W_a = \{j \in X_a : q_{a, j} > \widehat p_{a,j}\}$. Therefore, \Cref{lem:score-generator} gives
  \[\E[u_a(W_a)]\ge \left(\frac12-\eps\right)\E[u_a(X_a)]
+\E\!\left[\vq_a(X_a)\right]-\E\!\left[\sum_{j\in X_a}\widehat p_{a,j}\right].\]
Summing over $a\in\mathcal A$ gives
\begin{equation}\label{eq:three-sample-pooled-charge}
\E[H]\ge \left(\frac12-\eps\right)\E[T]
+\E\!\left[\sum_a\vq_a(X_a)\right]-\E\!\left[\sum_a\sum_{j\in X_a}\widehat p_{a,j}\right].
\end{equation}

For item $j$, let $S_j=\sum_{a:j\in X_a}\widehat p_{a,j}$ be the sum of the rival maxima, $M_j=\max_a q_{a,j}$ be the winning score, and $Q_j=\sum_{i=1}^n(q_{i,j}+q'_{i,j})$ be the total score. For now, assume $M_j > 0$. Each rival maximum is at most $M_j$, so $S_j\le d_jM_j$. Recall from \Cref{subsec:tentative} that $d_j=\sum_{i=1}^n(\mathbf{1}[j\in X_i]+\mathbf{1}[j\in X_i'])$ is the total tentative demand for item $j$. Since we also have $M_j\le Q_j\le d_jM_j$, we obtain
\[
S_j-Q_j
\le d_jQ_j-d_j(Q_j-M_j)-Q_j
\le d_jQ_j-\frac{Q_j^2}{M_j}.
\]
Summing the bounds in \Cref{lem:three-sample-competition} over the sample and real scores gives $\E[d_jQ_j]\le 5\E[Q_j]$, so
\[
\E[S_j-Q_j]\le\E\!\left[5Q_j-\frac{Q_j^2}{M_j}\right]
\le \frac{25}{4}\E[M_j],
\]
where the last inequality follows from $5x-x^2\le 25/4$ at $x=Q_j/M_j$. Note that this inequality is still true with $M_j = 0$, as both sides are now $0$.

Each item with a positive score contributes its winning score $M_j$ to its winner's bundle. Applying the subset bound from \Cref{lem:score-generator} to each $W_a\subseteq X_a$ gives
\[
\sum_{j\in [m]}M_j=\sum_{a\in\mathcal A}\vq_a(W_a)
\le \sum_{a\in\mathcal A}\bigl(u_a(W_a)+\eps u_a(X_a)\bigr)
=H+\eps T.
\]
Combining the bound on $\E[S_j-Q_j]$ with \eqref{eq:three-sample-pooled-charge} gives $\E[H]\ge (\frac12-\eps)\E[T]-\frac{25}{4}\E[H+\eps T]$, or $\E[H] \ge \left(\frac{2}{29} - \eps\right) \E[T]$ as claimed.
\end{proof}

\subsection{Welfare of the Algorithm}
\label{subsec:three-sample-rounding}

We are now ready to prove the final theorem. In particular, we show that the factor by which the algorithm's welfare falls short of $H$ is at most $4$. Intuitively, the algorithm gets the welfare from the highest scorer on item $j$ whenever the runner-up score belongs to a sample tuple and the highest score belongs to a real tuple. The symmetry between the sample and real valuations lets us apply subadditivity.

\begin{proof}[Proof of \Cref{thm:three-sample}]
Let us condition on the set of all $2n$ tentative tuples in $\mathcal{A}$. Then, for the two tuples belonging to each buyer $i$, we can use a fair coin to decide which tuple is the sample tentative tuple $(s_i, X_i, \vq_i)$ and which is the real tentative tuple $(r_i, X'_i, \vq'_i)$.

Fix a tuple $a$ from buyer $i$ and condition on it being a real tuple. For $j\in W_a$, the runner-up is the tuple with the second-highest positive score on $j$, if one exists. Let $G_a$ contain the items of $W_a$ for which the runner-up is a sample tuple or is absent. Note that for all items $j \in G_a$, $a$ is the only real tuple whose score exceeds the threshold, so buyer $i$'s final allocation $A_i$ contains $j$. In particular, $G_a \subseteq A_i$.

Keep the roles of buyer $i$'s tuples fixed and swap the roles of the two tuples for every other buyer. Define $G_a'$ by applying the rule for $G_a$ to the swapped roles. It contains the items of $W_a$ whose runner-up is now a sample tuple, as well as those with no runner-up. The independent fair coins for the other buyers make $G_a$ and $G_a'$ identically distributed conditional on $a$ being real.

The following cases show that $W_a=G_a\cup G_a'$.
\begin{itemize}
\item If the runner-up belongs to another buyer, this tuple is a sample tuple in exactly one of the two role assignments.
\item If the runner-up is buyer $i$'s other tuple, it is a sample tuple in both role assignments.
\item If there is no runner-up, the item belongs to both $G_a$ and $G'_a$.
\end{itemize}

Subadditivity then gives $u_a(W_a)\le u_a(G_a)+u_a(G_a')$. Since $G_a\subseteq A_i$ and $G_a$ and $G_a'$ have the same distribution conditional on $a$ being real, we have $\E[u_a(A_i)\mid a\text{ is real}]\ge u_a(W_a)/2$.

Since each tuple is real with probability $1/2$, summing over all tuples $a$ and combining with \Cref{lem:welfare-highest-score} and the identity for $T$ gives
\[
\E[\Alg]\ge \frac14\E[H]\ge \frac1{60}\E[T]
=\frac1{30}\E\!\left[\sum_{i=1}^n r_i\bigl(\Off_i(\vr)\bigr)\right].
\]
\end{proof}

\subsection{Extension to the Single-Sample Setting}\label{subsec:three-to-one-sample}

To obtain a single-sample online allocation algorithm, we apply the following black-box $k$-to-single-sample reduction from \cite{dutting2024online}, which increases the competitive ratio by a factor of $k + 1$.

\begin{theorem}[Theorem 3.10 of \cite{dutting2024online}]\label{thm:single-sample-reduction}
Suppose a $k$-sample online allocation algorithm in the Game of Googol obtains at least a $\gamma$ fraction of the expected offline optimum on every instance. Then there exists a single-sample online allocation algorithm in the Game of Googol that obtains at least a $\gamma/(k+1)$ fraction of the expected offline optimum on every instance.
\end{theorem}

\begin{proof}[Proof of \Cref{corr:ssalg}]
We draw a deterministic allocation algorithm from $\Off$ independently of the game and use it for all offline calls in \Cref{alg:three-sample}. \Cref{thm:three-sample} applies to each choice of deterministic allocator. Averaging over this choice and using that $\Off$ obtains at least an $\alpha$ fraction of the optimum in expectation on every profile, we have
\[
\E[\Alg]\ge \frac1{30}\E\!\left[\sum_{i=1}^n r_i\bigl(\Off_i(\vr)\bigr)\right]\ge \frac{\alpha}{30}\E_{\vr}[\Opt(\vr)].
\]
Applying \Cref{thm:single-sample-reduction} with $k=3$ and $\gamma=\alpha/30$ gives $\E[\SSAlg]\ge \frac{\alpha}{120}\E_{\vr}[\Opt(\vr)]$, as claimed.
\end{proof}

\section{Computing the Score Generator}
\label{sec:score-efficiency}

In this section, we discuss the computation of the score generator of \Cref{lem:score-generator}. For any score vector $\vq$ and any rival score vector $\vp$, define
\[G(\vq,\vp)=v\bigl(\{j\in X:q_j>p_j\}\bigr)-\vq(X)+\vp(X).\]
We seek a distribution over $\vq$ such that $\E[G(\vq,\vp)]\ge (1/2-\eps)V$ for every fixed $\vp$ and $\vq(S)\le v(S)+\eps V$ for every $S\subseteq X$ on each draw (recall that $V = v(X)$).

Following the proof of \Cref{lem:score-generator}, let $\delta=\eps V/|X|$ and $\mathcal G=\{0,\delta,\ldots,\lceil V/\delta\rceil\delta\}^{X}$. Let $\mathbf{1}_X$ be the indicator vector of $X$ and set $\mathcal Q=\{\vx+\delta\mathbf{1}_X:\vx\in\mathcal G,\ \vx(S)\le v(S)\text{ for all }S\subseteq X\}$, with vectors on $X$ extended by zero. We can then compute the score distribution by solving the following LP and proceeding as in the rest of the proof of \Cref{lem:score-generator}.
\[
\begin{aligned}
    \text{maximize}\quad & t \\
    \text{subject to}\quad & \sum_{\vq\in\mathcal Q}\mu_{\vq}G(\vq,\vp)\ge t && \text{for every }\vp\in\mathcal G,\\
    & \sum_{\vq\in\mathcal Q}\mu_{\vq}=1,\\
    & \mu_{\vq}\ge 0 && \text{for every }\vq\in\mathcal Q.
\end{aligned}
\]

However, this computation is generally inefficient. We show below that even with demand and value queries, one cannot efficiently find a score distribution that approximates $V$ to any fixed constant factor. On the positive side, we identify a subclass of subadditive valuations for which one can efficiently compute the score generator~--- and hence construct an efficient online combinatorial allocation algorithm with demand queries~--- while valuations in this subclass cannot be approximated by XOS valuations up to an $o(\log m)$ factor, matching the tight approximation gap established by \cite{bhawalkar2011welfare}.



\subsection{Oracle Lower Bounds}
\label{subsec:score-oracle-lower}

For simplicity, we fix the tentative bundle to be $X = [m]$. We prove the following proposition.

\begin{proposition}
\label{prop:score-oracle-lower}For any fixed $0<\alpha\le 1/2$, let $\RSG$ be an algorithm with value and demand query access to $v$ that outputs a nonnegative score vector $\vq$. Suppose $\E[G(\vq,\vp)]\ge\alpha V$ for every $v\in\mathcal V$ and every fixed rival score vector $\vp$, where $\vp$ is hidden from $\RSG$ and the expectation is over its randomness. Then $\RSG$ must make $\exp(\Omega_\alpha(m))$ queries in the worst case.
\end{proposition}

\begin{proof}
Fix a constant $M>1$. Set $f(0)=0$. For $0<x\le1$, define
\[
H(x)=\ln(16M/x),\qquad f(x)=\max\bigl\{1,M-\log_3(H(x)/H(1))\bigr\}.
\]
We first construct the reference valuation $v_0(S)=f(|S|/m)$, which is normalized and monotone with $v_0(X)=M$. To prove subadditivity, take nonempty bundles $S$ and $T$ with $|S|\ge |T|$, and set $x=|S\cup T|/m$ and $y=|S|/m$. Since $|S\cup T|\le |S|+|T|\le 2|S|$, we have $y\ge x/2\ge x^3/(16M)^2$, where the last inequality uses $x\le 1$ and $M>1$. Then $H(y)\le3H(x)$, so $v_0(S\cup T)-v_0(S)=f(x)-f(y)\le1$. Since $T$ is nonempty, $v_0(T)\ge 1$, and hence $v_0(S\cup T)\le v_0(S)+v_0(T)$. If either bundle is empty, subadditivity holds with equality because $v_0(\varnothing)=0$.

We then construct a family of subadditive valuations parameterized by $\vc \in \R_{\ge 0}^m$ as follows: for any $S \subseteq [m]$,
\[
v_{\vc}(S)=\min_{T\subseteq S}\left\{v_0(S\setminus T)+\mathbf{1}[T\ne \varnothing ]+\frac1m\sum_{j\in T}c_j\right\}.
\]

The valuation function above is monotone and subadditive for any $\vc$. To see subadditivity, fix $S$ and $S'$, and let $T$ and $T'$ be minimizers in the definition of $v_{\vc}(S)$ and $v_{\vc}(S')$, respectively. Then
\begin{align*}
    v_{\vc}(S\cup S')
    &\le v_0\bigl((S\cup S')\setminus(T\cup T')\bigr)+\mathbf{1}[T\cup T'\ne\varnothing]+\frac1m\sum_{j\in T\cup T'}c_j\\
    &\le v_0(S\setminus T)+v_0(S'\setminus T')+\mathbf{1}[T\ne\varnothing]+\mathbf{1}[T'\ne\varnothing]\\
    &\qquad+\frac1m\sum_{j\in T}c_j+\frac1m\sum_{j\in T'}c_j\\
    &=v_{\vc}(S)+v_{\vc}(S').
\end{align*}

We now show that any algorithm requires exponentially many queries by testing it on a random pair $(\vc,\vp)$: for each item $j$, draw $U_j\sim\operatorname{Unif}(0,1)$ independently and set $c_j=\tfrac12 U_j^{-3/2}$ and $p_j=(U_j^{-1/2}-1)/m$. We first claim that there is a bound $\delta_m=\exp(-\Omega_M(m))$ such that for every fixed $S\subseteq[m]$,
\begin{equation}\label{eq:same}
\Pr_{\vc}\bigl[v_{\vc}(S)\ne v_0(S)\bigr]\le\delta_m.
\end{equation}

Let $x=|S|/m$. If $f(x)\le 1$, then $T=\varnothing$ gives $v_0(S)=f(x)$ while every nonempty $T$ costs at least $1$, so $v_{\vc}(S)=v_0(S)$. Now suppose $f(x)>1$. Then $H(x)<3^{M-1}H(1)$, so $x>(16M)^{1-3^{M-1}}$. In particular, $x$ is bounded below by a positive constant depending only on $M$. Let $B=\{j\in S:U_j\le (x/(8M))^2\}$. Each $j\in S$ lies in $B$ independently with probability $(x/(8M))^2$, so $\E[|B|]=mx^3/(64M^2)$ and a Chernoff bound gives
\[
\Pr\bigl[|B|<\tfrac12\E[|B|]\bigr]\le\exp\bigl(-\E[|B|]/8\bigr)=\exp\!\left(-\frac{mx^3}{512M^2}\right)\le\delta_m.
\]
Consider when $|B|\ge\tfrac12\E[|B|]$. Every $j\in B$ has cost $c_j\ge\tfrac12(8M/x)^3=256M^3/x^3$, so any nonempty $T\subseteq S$ with $\sum_{j\in T}c_j<mM$ contains fewer than $mx^3/(256M^2)=\tfrac14\E[|B|]$ items of $B$. Hence $|S\setminus T|\ge|B\setminus T|\ge\tfrac14\E[|B|]=m\cdot x^3/(16M)^2$, which implies $H(|S\setminus T|/m)\le 3H(x)$ and thus $v_0(S\setminus T)+1\ge v_0(S)$. Every nonempty $T$ therefore costs at least $v_0(S)$ (either via $\tfrac1m\sum_{j\in T}c_j\ge M\ge v_0(S)$ or via $v_0(S\setminus T)+1\ge v_0(S)$), so $v_{\vc}(S)=v_0(S)$, proving \eqref{eq:same}.

We now bound the payoff of any fixed score vector $\vq$ independent of $(\vc,\vp)$. Let $W=\{j\in X:q_j>p_j\}$. Choosing $T=W$ in the definition of $v_{\vc}$ gives $v_{\vc}(W)\le 1+\frac1m\sum_{j\in W}c_j$ (which also holds if $W=\varnothing$). By the definition of $p_j$, we have $j\in W$ precisely when $U_j>(1+mq_j)^{-2}$, and hence
\[
\E\bigl[c_j\mathbf{1}[j\in W]\bigr]=\int_{(1+mq_j)^{-2}}^1 \tfrac12 u^{-3/2}\,du=mq_j.
\]
Summing over $j\in X$ yields $\E[v_{\vc}(W)]\le 1+\vq(X)$. Moreover, $\E[p_j]=(\E[U_j^{-1/2}]-1)/m=1/m$ for each item, so $\E[\vp(X)]=1$. Therefore,
\begin{equation}\label{eq:payoff-bound}
\E[G_{\vc}(\vq,\vp)]=\E[v_{\vc}(W)]-\vq(X)+\E[\vp(X)]\le 2.
\end{equation}

We finally bound the number of queries needed. Suppose $\RSG$ uses at most $Q$ queries. Run it on $v_0$ and $v_{\vc}$ with $X=[m]$ and the same random coins $R$, drawn independently of $(\vc,\vp)$, to obtain scores $\vq_0$ and $\vq_{\vc}$. Let $E$ be the event that the two query transcripts coincide. Conditional on $R$, the run on $v_0$ is fixed, so \eqref{eq:same} and a union bound over its queries give $\Pr[\overline{E} \mid R]\le Q\delta_m$.

On $E$, we have $\vq_{\vc}=\vq_0$, so the payoffs coincide. Outside $E$, let $W_{\vc}$ and $W_0$ be the sets won by $\vq_{\vc}$ and $\vq_0$, respectively, against $\vp$. Then
\[
G_{\vc}(\vq_{\vc},\vp)-G_{\vc}(\vq_0,\vp)
=v_{\vc}(W_{\vc})-v_{\vc}(W_0)-\vq_{\vc}(X)+\vq_0(X).
\]
Since $0\le v_{\vc}\le M$ and all scores are nonnegative, the two cases give
\[
G_{\vc}(\vq_{\vc},\vp)-G_{\vc}(\vq_0,\vp)\le \mathbf{1}[\overline{E}]\bigl(M+\vq_0(X)\bigr).
\]
Conditional on $R$, the score $\vq_0$ is fixed. Thus the conditional expectation of the right-hand side is $(M+\vq_0(X))\Pr[\overline{E}\mid R]$, which is at most $Q\delta_m(M+\vq_0(X))$. Averaging over $R$, we obtain
\begin{equation} \label{eq:payoff-diff}
\E\bigl[G_{\vc}(\vq_{\vc},\vp)-G_{\vc}(\vq_0,\vp)\bigr]\le Q\delta_m\bigl(M+\E[\vq_0(X)]\bigr)\le 2MQ\delta_m,
\end{equation}
where $\E[\vq_0(X)]\le M$ follows from the algorithm's guarantee on $v_0$ against zero rival prices ($\alpha M\le M-\E[\vq_0(X)]$).

Finally, combining $\RSG$'s guarantee on $v_{\vc}$ with \eqref{eq:same} applied to $X$ (which gives $\E[v_{\vc}(X)]\ge M(1-\delta_m)$), we have
\begin{align*}
    \alpha M(1-\delta_m) &\le\alpha\E[v_{\vc}(X)] \\
    &\le \E[G_{\vc}(\vq_{\vc},\vp)] \\
    &\le \E[G_{\vc}(\vq_0,\vp)]+2MQ\delta_m \\
    &\le 2 + 2MQ\delta_m
\end{align*}
where the last two inequalities follow from \eqref{eq:payoff-bound} and \eqref{eq:payoff-diff}, using the independence of $\vq_0$ and $(\vc, \vp)$.

Choosing $M=4/\alpha$ gives $2-4\delta_m\le 2MQ\delta_m$. For $m$ large enough that $\delta_m\le 1/4$, this implies $Q\ge (1-2\delta_m)/(M\delta_m)\ge 1/(2M\delta_m)=\exp(\Omega_\alpha(m))$.
\end{proof}

\subsection{Set-Cover Valuations with an Efficient Score Generator}
\label{subsec:penalized-cover-scores}

We now exhibit a family of subadditive valuations that admit an efficient random score generator. For this family, our reduction yields a constant-competitive single-sample prophet inequality in expected polynomial time given demand-query access. The family contains valuations with a tight \(\Theta(\log m)\) gap from XOS, so extending prior single-sample guarantees via XOS approximation would incur a logarithmic loss.

\paragraph{Set cover valuations with concave penalty.}
Let $m$ denote the number of items. A valuation in the family is parameterized by $k=O(\log m)$ and a concave nondecreasing function $\phi:\{0,\ldots,m\}\to\mathbb R_{\ge0}$ with $\phi(0)=0$. Let $T_1,\ldots,T_k\subseteq[m]$ be explicit sets with costs $c_1,\ldots,c_k>0$. For $D\subseteq[k]$, let $\vT(D)=\bigcup_{i\in D}T_i$ and $\vc(D)=\sum_{i\in D}c_i$. Define
\begin{equation}
  v(S)=\min_{D\subseteq[k]}
       \bigl\{\vc(D)+\phi(|S\setminus \vT(D)|)\bigr\}.
  \label{eq:penalized-cover-value}
\end{equation}
It is obvious that this valuation is monotone and has $v(\varnothing) = 0$. For subadditivity, consider two disjoint subsets $S,T \subseteq [m]$. Let $D_S,D_T$ be minimizing cover choices for $S,T$, with residual counts $r_S=|S\setminus\vT(D_S)|$ and $r_T=|T\setminus\vT(D_T)|$. Let $r=r_S+r_T\le m$, and note that concavity and the identity $\phi(0)=0$ give $\phi(r_S)+\phi(r_T) \ge \phi(r)$. Finally, suppose we cover $S \cup T$ with $D_S\cup D_T$. Every item of $S\cup T$ that it leaves uncovered was already uncovered by the choice for its own bundle, so at most $r$ items remain uncovered. Therefore,
\begin{align*}
v(S\cup T)
&\le \vc(D_S\cup D_T)+\phi\bigl(|(S\cup T)\setminus\vT(D_S\cup D_T)|\bigr)\\
&\le \vc(D_S)+\vc(D_T)+\phi(r)\\
&\le \bigl(\vc(D_S)+\phi(r_S)\bigr)+\bigl(\vc(D_T)+\phi(r_T)\bigr)
=v(S)+v(T).
\end{align*}

Throughout, $k=O(\log m)$, so $v(S)$ can be evaluated by enumerating the
$2^k$ choices of $D$. We assume that $\phi$ is efficiently evaluable
and count arithmetic operations, comparisons, and uniform draws at unit
cost. For rational costs, penalty values, and target approximation,
the LP below also has polynomial bit complexity in the encoded input
and the size of the score grid.

\paragraph{The gap from XOS.} Set cover cost functions are a classical example of subadditive valuations with a tight $\Theta(\log m)$ gap from XOS. We show that this gap persists when uncovered items incur a penalty, provided the penalty function $\phi$ grows sufficiently large relative to the cost of a cover.
For a bundle $T$, let $A_v(T)$ be the maximum of $\va(T)$ over nonnegative
additive vectors $\va$ satisfying $\va(S)\le v(S)$ for every $S\subseteq[m]$.
The pointwise maximum of these additive functions is the largest XOS lower
bound on $v$. Thus the XOS approximation factor is
$\Gamma(v)=\max_{T:v(T)>0}v(T)/A_v(T)$, with $\Gamma(v)=1$ for
the zero valuation.

Consider the following set system of Bhawalkar and Roughgarden~\cite{bhawalkar2011welfare}. Let $k$ be a multiple of four, and let each of the $m=\binom{k}{k/2}$ items encode a $k/2$-subset of $[k]$. Let $T_i$ consist of the items whose subsets contain $i$, and give every cover the same cost $C>0$. Buying $d$ covers leaves exactly $N_d=\binom{k-d}{k/2}$ items uncovered, with $N_{k/2+1}=0$. Hence
\begin{equation}\label{eq:penalized-cover-grand}
 v([m])=\min_{0\le d\le k/2+1}\{Cd+\phi(N_d)\}.
\end{equation}

On the other hand, each item belongs to $k/2$ covers, so every additive lower bound satisfies $(k/2)\va([m])=\sum_i \va(T_i)\le kC$. The empty cover also gives $\va([m])\le\phi(m)$, so
\[A_v([m]) \le \min\{2C, \phi(m)\}.\]

Combining the two, and noting that $v([m]) \ge \min\{C, \phi(m)\}$ by choosing $d = 1$ in \eqref{eq:penalized-cover-grand} so $A_v([m]) \le 2 v([m])$, we have $\Gamma(v)\ge \frac{1}{2} \max\left\{1, \frac{v([m])}{C}\right\}$.
This gives two criteria as $k\to\infty$. Here $C$ and $\phi$ may vary with $k$.
\begin{itemize}
\item \emph{Logarithmic gap.} The gap is $\Theta(\log m)=\Theta(k)$ if for some fixed $\delta\in(0,1/2)$,
\begin{equation}
             \phi(N_{\lfloor\delta k\rfloor})=\Omega(kC).
             \label{eq:penalized-cover-log-gap}
\end{equation}
Indeed, buying at least $\lfloor\delta k\rfloor$ covers costs $\Omega(kC)$, and buying fewer leaves at least $N_{\lfloor\delta k\rfloor}$ residual items and incurs a penalty of $\Omega(kC)$. Thus $v([m])=\Omega(kC)$, and $A_v([m])\le 2C$ gives $\Gamma(v)=\Omega(\log m)$. The general $O(\log m)$ XOS approximation for subadditive valuations gives the matching upper bound.
\item \emph{Superconstant gap.} The gap tends to infinity if $\phi(m)/C\to\infty$. For any fixed integer $d\ge 1$, we have $N_d/m\to 2^{-d}$, so concavity gives $\phi(N_d)\ge (N_d/m)\phi(m)=\omega(C)$. Hence \eqref{eq:penalized-cover-grand} gives $v([m])\ge \min\{dC,\phi(N_d)\}\ge dC$ for all sufficiently large $k$. Since $d$ is arbitrary and $A_v([m])\le 2C$, the gap tends to infinity.
\end{itemize}

\paragraph{An LP for the score marginals.}
We return to arbitrary explicit covers and costs. For any tentative bundle $X$, we relax the LP at the beginning of this section by keeping only the marginal distribution of each item’s score. For each cover choice $D$, these marginals determine the expected number of uncovered winning items. Applying $\phi$ to this expectation upper-bounds the expected penalty by concavity. There are only $2^k=\mathrm{poly}(m)$ cover choices, and the interpolated penalty is the minimum of at most $m$ affine functions. For each cover choice and affine function, we can minimize over rival prices one item at a time. The resulting LP has polynomial size, and under the condition below, rounding its solution yields a score generator satisfying Lemma~\ref{lem:score-generator}.

\begin{lemma}
\label{lem:penalized-cover-scores}
Fix $0<\eps<1/2$ and let $c_{\min}=\min_i c_i$. If
\begin{equation}
       c_{\min} \ge \frac{16\,\phi(1)\ln(k+1)}{\eps^2},
       \label{eq:penalized-cover-score-condition}
\end{equation}
then for every tentative bundle $X$ there is a polynomial-time algorithm that draws a score vector $\vq$ supported on $X$ such that $\mathbb E[G(\vq,\vp)]\ge(1/2-\eps)v(X)$ for every fixed $\vp\ge0$. On every draw, $\vq(S)\le v(S)$ for all $S\subseteq X$.
\end{lemma}

\begin{proof}
Let $V=v(X)$ and $n=|X|$. Recall that the payoff is $G(\vq,\vp)=v(\{j\in X:q_j>p_j\})-\vq(X)+\vp(X)$. If $V=0$, return zero. If $0<V<c_{\min}$, every $S\subseteq X$ has $v(S)=\phi(|S|)$, since buying even one cover costs more than $v(S)$. The uniform weight vector $\va$ on $X$ of total $V$ is therefore an additive lower bound on $X$. Return $\vq=U\va$ for a uniform $U\in[0,1]$. For a fixed price vector $\vp$, the expected winning value is at least $\sum_{j\in X} a_j \Pr[q_j\ge p_j] = \sum_{j\in X}(a_j-p_j)_+\ge V-\vp(X)$, and $\mathbb E[\vq(X)]=V/2$. This gives payoff at least $V-\vp(X) -V/2+\vp(X) = V/2$ and the claimed subset bound. Henceforth suppose that $V\ge c_{\min}$.

\smallskip\noindent\emph{Computing the marginals.}
For $D\subseteq[k]$, let $R_D=X\setminus \vT(D)$. Extend $\phi$ to $[0,n]$ by interpolating it linearly between consecutive integers. Write the interpolated penalty as $\phi(r)=\min_{1\le\ell\le n}\{a_\ell r+b_\ell\}$. For $\vz\in[0,1]^X$, set
\[
       F(\vz)=\min_{D\subseteq[k]}\{\vc(D)+\phi(\vz(R_D))\}.
\]
For any random set $W$ with inclusion probabilities $z_j$, we have $\mathbb E[v(W)]\le F(\vz)$. Indeed, for each fixed $D$, taking expectations in \eqref{eq:penalized-cover-value} and applying Jensen's inequality bounds $\mathbb E[v(W)]$ by $\vc(D)+\phi(\vz(R_D))$.

Let $\delta=\eps V/(2n)$ and use the grid $\mathcal{G}=\delta\{0,1,\ldots,\lceil V/\delta\rceil+1\}$. The LP variables $x_{j,b}$ give the marginal probability of bidding $b\in\mathcal{G}$ on $j\in X$. Let
\[
 z_j(s)=\sum_{b>s}x_{j,b},\qquad
 L(\vx)=\sum_{j\in X}\sum_{b\in\mathcal{G}}b x_{j,b}.
\]
We seek marginals $\vx$ satisfying
\begin{equation}
 F\bigl((z_j(p_j))_{j\in X}\bigr)-L(\vx)+\vp(X)
     \ \ge\ (1/2-\eps/2)V\qquad\text{for every }\vp\ge0.
 \label{eq:penalized-cover-relaxation}
\end{equation}
The finite-grid construction in the proof of Lemma~\ref{lem:score-generator}, with precision $\eps/2$, gives a joint score distribution on $\mathcal{G}^X$ with this payoff for the actual winning value. The extra grid point accounts for its shift by $\delta$ to win ties strictly. The marginals of this joint distribution satisfy \eqref{eq:penalized-cover-relaxation} by the preceding Jensen bound.

We now write the LP explicitly. For each $j\in R_D$, the rival minimizes $s+a_\ell z_j(s)$, and it sets $s=0$ outside $R_D$. It suffices to consider $s\in\mathcal{G}$, since $z_j(s)$ is constant from one grid point up to the next and is zero beyond the last grid point. We can therefore enforce \eqref{eq:penalized-cover-relaxation} with the following linear constraints.
\begin{align*}
 &x_{j,b}\ge0,\qquad \sum_{b\in\mathcal{G}}x_{j,b}=1
       &&(j\in X),\\
 &y_{j,\ell}\le s+a_\ell z_j(s)
       &&(j\in X,\ 1\le\ell\le n,\ s\in\mathcal{G}),\\
 &\vc(D)+b_\ell+\sum_{j\in R_D}y_{j,\ell}-L(\vx)\ge(1/2-\eps/2)V
       &&(D\subseteq[k],\ 1\le\ell\le n).
\end{align*}
Indeed, each $y_{j,\ell}$ can be set to the minimum over $s$ in the second line. There are $O(n/\eps)$ grid points and $2^k$ cover choices, so the LP has polynomial size. Take any feasible solution and draw the scores independently using its marginals.

\smallskip\noindent\emph{The loss from independent draws.}
Fix $\vp$, let $W$ be the resulting winning set, and let $H=F((z_j(p_j))_{j\in X})\le V$. Only choices $D$ with $\vc(D)<H$ can give a realized cover cost below $H$. For each such $D$, let
\[
 u_D=H-\vc(D),\qquad Z_D=|W\cap R_D|,\qquad \mu_D=\mathbb E[Z_D].
\]
We have $0<u_D\le\phi(\mu_D)$ and thus $\mu_D>0$. Concavity gives
$\phi(r)\ge u_D\min\{r/\mu_D,1\}$. The shortfall below $H$ for this
cover choice is therefore at most the positive part of
$Y_D=u_D(1-Z_D/\mu_D)$. Since $Z_D$ is a sum of independent Bernoulli
variables, for every $\lambda>0$,
\[
 \mathbb E[e^{\lambda Y_D}]
 \le \exp\!\left(\lambda u_D+
       \mu_D(e^{-\lambda u_D/\mu_D}-1)\right)
 \le \exp\!\left(\frac{\lambda^2u_D^2}{2\mu_D}\right).
\]
The last inequality uses $e^{-t}\le1-t+t^2/2$ for $t\ge0$.

Let $N$ be the number of choices with $\vc(D)<H$ and let $\sigma^2$ be an upper bound on $u_D^2/\mu_D$ for these choices. The overall shortfall is at most $\max\{0,\max_D Y_D\}$, where the inner maximum ranges over these $N$ choices. For $N>0$, bounding the exponential of this maximum by $1+\sum_D e^{\lambda Y_D}$, applying Jensen's inequality, and minimizing over $\lambda$ gives
\[
 H-\mathbb E[v(W)]\le \sigma\sqrt{2\ln(N+1)}.
\]
For $N=0$ the shortfall is zero. Each relevant choice uses at most $\lfloor V/c_{\min}\rfloor$ covers. Encoding its sorted indices as a list padded with zeros gives $N\le(k+1)^{\lfloor V/c_{\min}\rfloor}$. Since $V\ge c_{\min}$ and $k\ge1$, $\ln(N+1)\le 2(V/c_{\min})\ln(k+1)$.

Finally, $u_D\le V$ and $u_D\le\phi(\mu_D)\le\phi(1)\mu_D$ give
$\sigma^2=V\phi(1)$ as a valid bound. Consequently,
\[
 H-\mathbb E[v(W)]
 \le 2V\sqrt{\frac{\phi(1)\ln(k+1)}{c_{\min}}}
 \le \frac{\eps V}{2}
\]
under \eqref{eq:penalized-cover-score-condition}. 
Combining this bound with \eqref{eq:penalized-cover-relaxation} gives $\mathbb E[G(\vq,\vp)]\ge(1/2-\eps)V$.

\smallskip\noindent\emph{Enforcing the subset bound.}
This step follows the proof of \Cref{lem:score-generator}. If a draw violates $\vq(S)\le v(S)$, set its scores on $S$ to zero. For every $\vp$, this saves $\vq(S)$ and reduces the winning value by at most $v(S)$, by subadditivity. The correction therefore increases the payoff pointwise. To find a violation, check $\vq(T_i\cap X)\le c_i$ for every $i$, and check that the sum of the $r$ largest scores is at most $\phi(r)$ for every $1\le r\le n$. A failed check exhibits a violating subset. If all checks pass, then for every $S\subseteq X$ and $D\subseteq[k]$ the covered items in $S$ have weight at most $\vc(D)$ and the remaining items have weight at most $\phi(|S\setminus \vT(D)|)$. Minimizing over $D$ proves $\vq(S)\le v(S)$. Repeat until all checks pass. Each correction removes at least one positive score, so at most $n$ corrections are needed.
\end{proof}

\paragraph{When both guarantees hold.} For an explicit example satisfying Equations \eqref{eq:penalized-cover-log-gap} and \eqref{eq:penalized-cover-score-condition}, take $c_i=C$ for all $i\in [k]$ and $\phi(s)=s/M$ for any $M$ and $C$ satisfying $16\ln(k+1)/\eps^2 \le MC \le m^{1-\gamma}$ for some fixed $\gamma>0$. The lower bound gives \eqref{eq:penalized-cover-score-condition}, since $\phi(1)=1/M$. For the gap, choose a sufficiently small constant $\delta>0$ so that $N_{\lfloor\delta k\rfloor}\ge m^{1-\gamma/2}$. Since $k=\Theta(\log m)$, this gives $N_{\lfloor\delta k\rfloor}=\Omega(kMC)$, as required by \eqref{eq:penalized-cover-log-gap}. Thus these valuations admit an efficient score generator with payoff at least $(1/2-\eps)v(X)$ and exhibit a tight $\Theta(\log m)$ gap from XOS.









\newpage
\bibliographystyle{alpha}
\bibliography{refs}

@inproceedings{AKW14,
  author    = {Pablo Daniel Azar and Robert Kleinberg and S. Matthew Weinberg},
  title     = {Prophet Inequalities with Limited Information},
  booktitle = {Proceedings of the Twenty-Fifth Annual ACM-SIAM Symposium on Discrete Algorithms},
  pages     = {1358--1377},
  year      = {2014},
  publisher = {SIAM},
  doi       = {10.1137/1.9781611973402.100}
}

@inproceedings{CDF22,
  author    = {Constantine Caramanis and Paul D{\"u}tting and Matthew Faw and
               Federico Fusco and Philip Lazos and Stefano Leonardi and
               Orestis Papadigenopoulos and Emmanouil Pountourakis and
               Rebecca Reiffenh{\"a}user},
  title     = {Single-Sample Prophet Inequalities via Greedy-Ordered Selection},
  booktitle = {Proceedings of the 2022 Annual ACM-SIAM Symposium on Discrete Algorithms},
  pages     = {1298--1325},
  year      = {2022},
  publisher = {SIAM},
  doi       = {10.1137/1.9781611977073.54}
}

@inproceedings{CCES20,
  author    = {Jos{\'e} R. Correa and Andr{\'e}s Cristi and Boris Epstein and Jos{\'e} A. Soto},
  title     = {The Two-Sided {Game of Googol} and Sample-Based Prophet Inequalities},
  booktitle = {Proceedings of the Thirty-First Annual ACM-SIAM Symposium on Discrete Algorithms},
  pages     = {2066--2081},
  year      = {2020},
  publisher = {SIAM},
  doi       = {10.1137/1.9781611975994.127}
}

@inproceedings{FGL15,
  author    = {Michal Feldman and Nick Gravin and Brendan Lucier},
  title     = {Combinatorial Auctions via Posted Prices},
  booktitle = {Proceedings of the Twenty-Sixth Annual ACM-SIAM Symposium on Discrete Algorithms},
  pages     = {123--135},
  year      = {2015},
  publisher = {SIAM},
  doi       = {10.1137/1.9781611973730.10}
}

@misc{FSW26,
  author        = {Matthew Faw and Sahil Singla and Yifan Wang},
  title         = {Online Allocation using Few Samples},
  year          = {2026},
  eprint        = {2609.17343},
  archivePrefix = {arXiv},
  primaryClass  = {cs.DS},
  note          = {To appear in SODA 2027},
  url           = {https://arxiv.org/abs/2609.17343}
}

@inproceedings{KW12,
  author    = {Robert Kleinberg and S. Matthew Weinberg},
  title     = {Matroid Prophet Inequalities},
  booktitle = {Proceedings of the Forty-Fourth Annual ACM Symposium on Theory of Computing},
  pages     = {123--135},
  year      = {2012},
  publisher = {ACM},
  doi       = {10.1145/2213977.2213991}
}

@inproceedings{RS17,
  author    = {Aviad Rubinstein and Sahil Singla},
  title     = {Combinatorial Prophet Inequalities},
  booktitle = {Proceedings of the Twenty-Eighth Annual ACM-SIAM Symposium on Discrete Algorithms},
  pages     = {1671--1687},
  year      = {2017},
  publisher = {SIAM},
  doi       = {10.1137/1.9781611974782.110}
}

@article{DFKL20,
  author    = {Paul D{\"u}tting and Michal Feldman and Thomas Kesselheim and Brendan Lucier},
  title     = {Prophet Inequalities Made Easy: Stochastic Optimization by Pricing Nonstochastic Inputs},
  journal   = {SIAM Journal on Computing},
  volume    = {49},
  number    = {3},
  pages     = {540--582},
  year      = {2020},
  doi       = {10.1137/20M1323850},
  note      = {Preliminary version in FOCS 2017}
}

@inproceedings{Zha20,
  author    = {Hanrui Zhang},
  title     = {Improved Prophet Inequalities for Combinatorial Welfare Maximization with (Approximately) Subadditive Agents},
  booktitle = {28th Annual European Symposium on Algorithms (ESA 2020)},
  series    = {Leibniz International Proceedings in Informatics (LIPIcs)},
  volume    = {173},
  pages     = {82:1--82:17},
  year      = {2020},
  publisher = {Schloss Dagstuhl--Leibniz-Zentrum f{\"u}r Informatik},
  doi       = {10.4230/LIPIcs.ESA.2020.82}
}

@inproceedings{KNR22,
  author    = {Haim Kaplan and David Naori and Danny Raz},
  title     = {Online Weighted Matching with a Sample},
  booktitle = {Proceedings of the 2022 Annual ACM-SIAM Symposium on Discrete Algorithms},
  pages     = {1247--1272},
  year      = {2022},
  publisher = {SIAM},
  doi       = {10.1137/1.9781611977073.52}
}

@misc{FuEtAl24,
  author        = {Hu Fu and Pinyan Lu and Zhihao Gavin Tang and Hongxun Wu and Jinzhao Wu and Qianfan Zhang},
  title         = {Sample-Based Matroid Prophet Inequalities},
  year          = {2024},
  eprint        = {2406.12799},
  archivePrefix = {arXiv},
  primaryClass  = {cs.DS},
  url           = {https://arxiv.org/abs/2406.12799}
}

@inproceedings{GSW25,
  author    = {Rohan Ghuge and Sahil Singla and Yifan Wang},
  title     = {Single-Sample and Robust Online Resource Allocation},
  booktitle = {Proceedings of the 57th Annual ACM Symposium on Theory of Computing},
  pages     = {1442--1453},
  year      = {2025},
  publisher = {ACM},
  doi       = {10.1145/3717823.3718246}
}

@misc{RS26,
  author        = {Aviad Rubinstein and Sahil Singla},
  title         = {Improved Combinatorial Auctions with and without Samples},
  year          = {2026},
  eprint        = {2610.17517},
  archivePrefix = {arXiv},
  primaryClass  = {cs.GT},
  note          = {To appear in SODA 2027},
  url           = {https://arxiv.org/abs/2610.17517}
}

@inproceedings{dutting2020prophet,
  author    = {Paul D{\"u}tting and Thomas Kesselheim and Brendan Lucier},
  title     = {An {$O(\log\log m)$} Prophet Inequality for Subadditive Combinatorial Auctions},
  booktitle = {2020 IEEE 61st Annual Symposium on Foundations of Computer Science (FOCS)},
  pages     = {306--317},
  year      = {2020},
  publisher = {IEEE},
  eprint    = {2004.09784},
  archivePrefix = {arXiv},
  url       = {https://arxiv.org/abs/2004.09784}
}

@unpublished{KMRS26,
  author = {T. Kroll and C. MacRury and R. Reiffenh{\"a}user
            and S. M. Sundaram},
  title  = {An efficient {$1/4$}-competitive prophet inequality
            for subadditive combinatorial allocations},
  year   = {2026},
  note   = {Submitted to {STOC} 2027, submission no.~1456}
}

@inproceedings{FFGL13,
  author    = {Michal Feldman and Hu Fu and Nick Gravin and Brendan Lucier},
  title     = {Simultaneous Auctions Are (Almost) Efficient},
  booktitle = {Proceedings of the Forty-Fifth Annual ACM Symposium
               on Theory of Computing},
  series    = {STOC '13},
  pages     = {201--209},
  publisher = {ACM},
  year      = {2013},
  doi       = {10.1145/2488608.2488634},
  url       = {https://doi.org/10.1145/2488608.2488634}
}

@inproceedings{correa2023constant,
  title={A constant factor prophet inequality for online combinatorial auctions},
  author={Correa, Jos{\'e} and Cristi, Andr{\'e}s},
  booktitle={Proceedings of the 55th Annual ACM Symposium on Theory of Computing},
  pages={686--697},
  year={2023}
}

@inproceedings{macrury2026efficient,
  title={Efficient Algorithms for Online Subadditive Combinatorial Allocations},
  author={MacRury, Calum and Neogi, Rian and Pashkovich, Kanstantsin and Singla, Sahil and Sundaram, Siddharth M. and Swamy, Chaitanya},
  booktitle={Proceedings of the 2027 Annual ACM-SIAM Symposium on Discrete Algorithms (SODA)},
  year={2027},
  note={To appear}
}

@article{chawla2026single,
  title={Single-Sample Prophet Inequalities: A Combinatorial to Single-Item Reduction},
  author={Chawla, Shuchi and Dang, Trung},
  journal={arXiv preprint arXiv:2610.03660},
  year={2026}
}

@inproceedings{rubinstein20samples,
  author    = {Rubinstein, Aviad and Wang, Jack Z. and Weinberg, S. Matthew},
  title     = {{Optimal Single-Choice Prophet Inequalities from Samples}},
  booktitle = {11th Innovations in Theoretical Computer Science Conference (ITCS 2020)},
  pages     = {60:1--60:10},
  isbn      = {978-3-95977-134-4},
  issn      = {1868-8969},
  year      = {2020},
  volume    = {151},
  url       = {https://drops.dagstuhl.de/entities/document/10.4230/LIPIcs.ITCS.2020.60},
  urn       = {urn:nbn:de:0030-drops-117452},
  doi       = {10.4230/LIPIcs.ITCS.2020.60},
  series    = {Leibniz International Proceedings in Informatics (LIPIcs)}
}

@inproceedings{dutting2024online,
  title         = {Online combinatorial allocations and auctions with few samples},
  author        = {D{\"u}tting, Paul and Kesselheim, Thomas and Lucier, Brendan and Reiffenh{\"a}user, Rebecca and Singla, Sahil},
  booktitle     = {2024 IEEE 65th Annual Symposium on Foundations of Computer Science (FOCS)},
  pages         = {1231--1250},
  year          = {2024},
  organization  = {IEEE},
  eprint        = {2409.11091},
  archiveprefix = {arXiv}
}

@article{feige2009maximizing,
  title={On Maximizing Welfare When Utility Functions Are Subadditive},
  author={Feige, Uriel},
  journal={SIAM Journal on Computing},
  volume={39},
  number={1},
  pages={122--142},
  year={2009},
  doi={10.1137/070680977}
}

@inproceedings{bhawalkar2011welfare,
  title={Welfare Guarantees for Combinatorial Auctions with Item Bidding},
  author={Bhawalkar, Kshipra and Roughgarden, Tim},
  booktitle={Proceedings of the 2011 Annual ACM-SIAM Symposium on Discrete Algorithms (SODA)},
  pages={700--709},
  year={2011},
  doi={10.1137/1.9781611973082.55}
}

@article{cristi2026optimal,
  title={The Optimal Competitive Ratio for Subadditive Online Combinatorial Auctions is Less Than 1/2},
  author={Cristi, Andr{\'e}s and Zhang, Jiechen},
  journal={arXiv preprint arXiv:2611.24920},
  year={2026}
}
\end{document}